\documentclass[conference]{IEEEtran}
\IEEEoverridecommandlockouts
\usepackage{cite}
\usepackage{amsmath,amssymb,amsfonts}
\usepackage{algorithmic}
\usepackage{graphicx}
\usepackage{textcomp}
\usepackage{xcolor}
\usepackage{amsmath}
\usepackage{amssymb}
\usepackage{float}

\def\BibTeX{{\rm B\kern-.05em{\sc i\kern-.025em b}\kern-.08em
    T\kern-.1667em\lower.7ex\hbox{E}\kern-.125emX}}
\begin{document}

\title{Future Mining: Learning for Safety and Security\\}

\author{\IEEEauthorblockN{
 Md Sazedur Rahman, Mizanur Rahman Jewel, Sanjay Madria}  
      \IEEEauthorblockA{%
 Department of Computer Science, Missouri University of Science and Technology, Rolla, MO 65401, USA}
      
\thanks{\IEEEauthorrefmark{4}This work was supported by a grant from CDC-NIOSH.} 
Emails: \texttt{\{mrvfw, mj9vc, madrias\}@mst.edu}\vspace{-0.15cm}}

\maketitle

\begin{abstract}
Mining industry is rapidly transforming into an AI-driven cyber-physical ecosystem where safety and operational reliability depend on robust perception, resilient communication, trustworthy distributed intelligence and continuous monitoring of miners and equipment. Real-world mining environments impose severe constraints like poor illumination, dust, occlusion, GPS-denied conditions, irregular underground topologies, and intermittent connectivity. These factors degrade perception quality, disrupt situational awareness, impair trajectory prediction and weaken the reliability of distributed learning systems. Emerging cyber-physical threats, including backdoor triggers, sensor spoofing, label-flip attacks and poisoned model updates, further jeopardize operational safety, particularly as mines increasingly adopt autonomous vehicles, humanoid assistance, and federated learning for collaborative intelligence. Moreover, energy-constrained sensors experience uneven and unpredictable battery depletion, creating blind spots in safety coverage and disrupting hazard detection pipelines. This paper presents a vision for a \textit{Unified Smart Safety and Security Architecture} that integrates multimodal perception, spatial-temporal modeling, secure federated learning, reinforcement learning, DTN-enabled communication and energy-aware sensing into a cohesive safety fabric. We detail five core modules: Miner-finder, Multimodal Situational Awareness, Backdoor Attack Monitor, TrustFED-LFD and IoT-driven Equipment Health Monitoring, addressing critical gaps in miner localization, hazard understanding, model integrity, federated robustness and predictive maintenance. Together, these modules form an end-to-end framework capable of detecting hazards, responding to disasters, guiding miners through obstructed pathways, identifying compromised models or sensors and ensuring the health of mission-critical equipment. By unifying these components, this work outlines a comprehensive research vision for building a futuristic,  resilient, proactive and trustworthy intelligent mining system capable of safeguarding miners and maintaining operational continuity under extreme and adversarial conditions.
\end{abstract}


\section{Introduction}
Mining operations are evolving complex cyber-physical systems driven by advances in sensing, connectivity and machine learning \cite{rojas2025ai, zhao2025open}. Modern underground and surface mines deploy heterogeneous sensors, including thermal and RGB cameras, LiDAR units, gas and pressure detectors, physiological monitors and high-resolution video surveillance, enabling real-time situational awareness for tasks such as miner trajectory prediction, fatigue assessment, crack detection in pipes, tunnel segmentation and machine anomaly detection. Recent advances in spatial-temporal transformers and large sequence models \cite{liu2024st} further strengthen these capabilities by enabling long-range mobility prediction and hazard anticipation. Yet mining environments pose severe constraints for continuous data exchange. Underground tunnels lack GPS and suffer from extreme signal attenuation, leaky-feeder limitations, and frequent network partitioning. Thus, traditional communication architectures fail to guarantee end-to-end connectivity. Delay-Tolerant Networking (DTN) becomes essential by enabling store-carry-forward transfer of safety-critical information across miners, humanoids, and mobile equipment, ensuring that hazard alerts, routing updates, and model synchronization continue even under intermittent connectivity \cite{goyal2024minerrouter}. 

Harsh sensing conditions further strain perception systems. Low-light scenarios, dust, and occlusion significantly degrade visual model performance. The DIS-Mine framework \cite{jewel2024dis} shows that standard segmentation models fail in dark underground conditions without specialized enhancement. Sensor reliability also becomes a major challenge. As shown in the CAV-AD framework \cite{rahman2024cav}, subtle manipulations in sensor readings can cause autonomous mining vehicles to behave unpredictably, which can lead to collisions or severe accidents. Distributed learning introduces additional vulnerabilities. Federated Learning (FL) across mines suffers from non-IID distributions, malicious clients, and low-quality data \cite{sun2024client}. The MineDetect framework \cite{rahman2025detecting} shows that untargeted attacks like sign-flip or additive-noise and noisy data generated by unreliable clients can severely degrade global model performance without robust aggregation. Reinforcement Learning (RL) based navigation frameworks, such as OGLe-Mine \cite{goyal2025ogle} demonstrate how post-disaster escape routing must adapt to obstacles, near-dark conditions, and dynamic layouts. Beyond these well-studied threats, a new class of \textit{machine unlearning attacks} has emerged where adversaries abuse unlearning procedures to erase influential safety-related samples, weaken decision boundaries, or indirectly inject bias into global models \cite{liu2024backdoor, liu2025threats}. 

\begin{figure}[!t]
    \centering
   \includegraphics[width=\linewidth,height=5cm]
{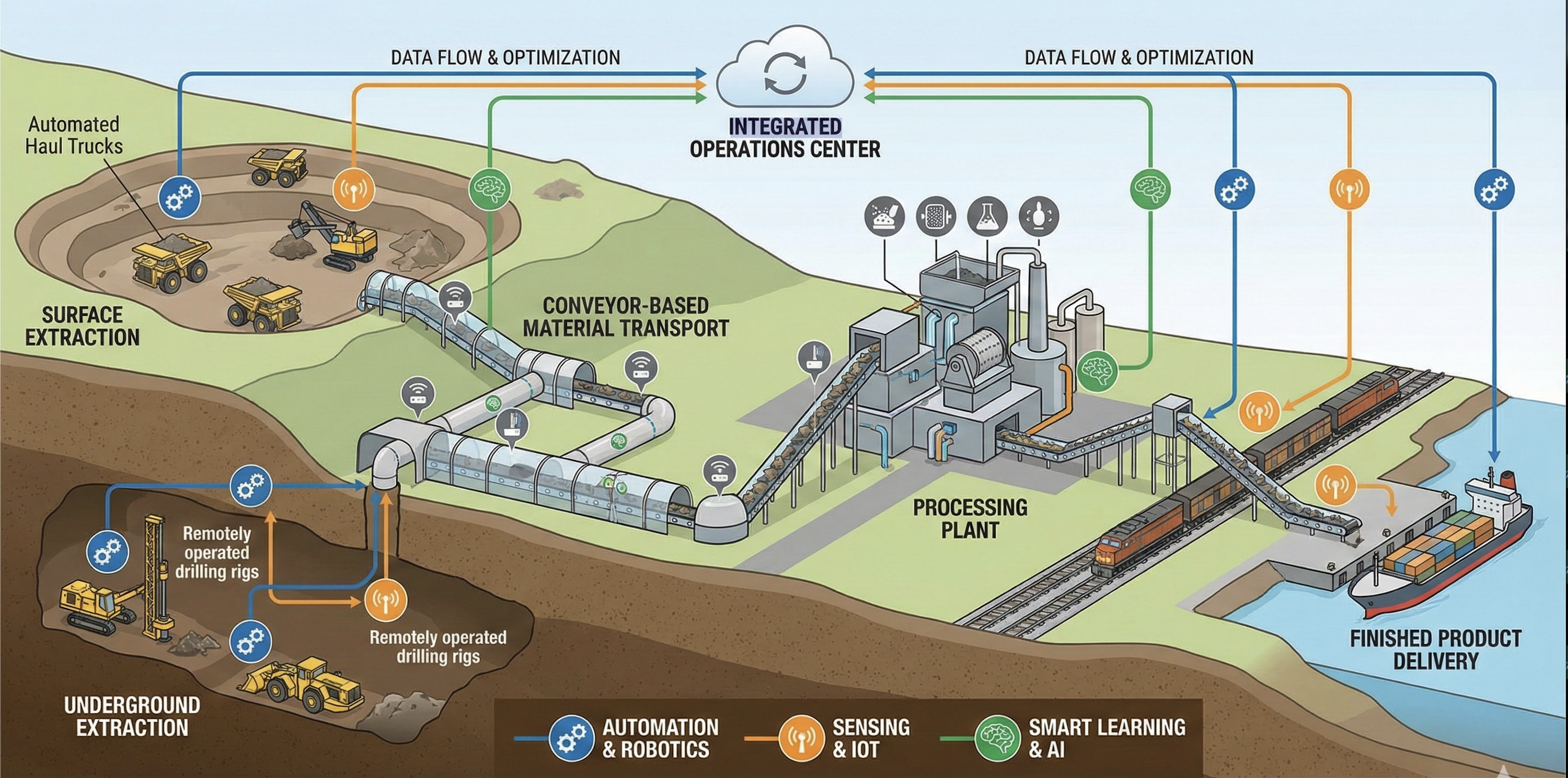}
    \caption{An illustrative overview of modern mining operations, including underground and surface extraction, heavy equipment coordination, conveyor-based material transport, processing plants, and finished product delivery. The figure highlights how sensing, automation, and smart learning can operate across the entire mining ecosystem [AI Generated].}
  \label{fig:smart_mine}
\end{figure}

Achieving reliable and secure smart mining requires a unified framework that integrates advanced computer vision, robust wireless communication, cybersecurity mechanisms, and distributed machine learning strategies \cite{rahman2025detecting}. To contextualize this vision, Fig~\ref{fig:smart_mine} illustrates the interconnected workflow of modern mining operations, spanning underground extraction, surface excavation, heavy machinery coordination, conveyor-based material transport, ore processing, and final product delivery. This multi-stage ecosystem highlights both the complexity and the opportunities for learning to enhance safety, efficiency, and security at every step.
We present a holistic vision that unifies advances in multimodal perception, spatial-temporal learning, secure FL or RL frameworks, and DTN-enabled communication, outlining how these complementary technologies can converge to enable resilient, trustworthy, and situation-aware smart mining ecosystems of the future.

\begin{figure*}[!ht]
    \centering
    \includegraphics[width=11cm,height=14cm]{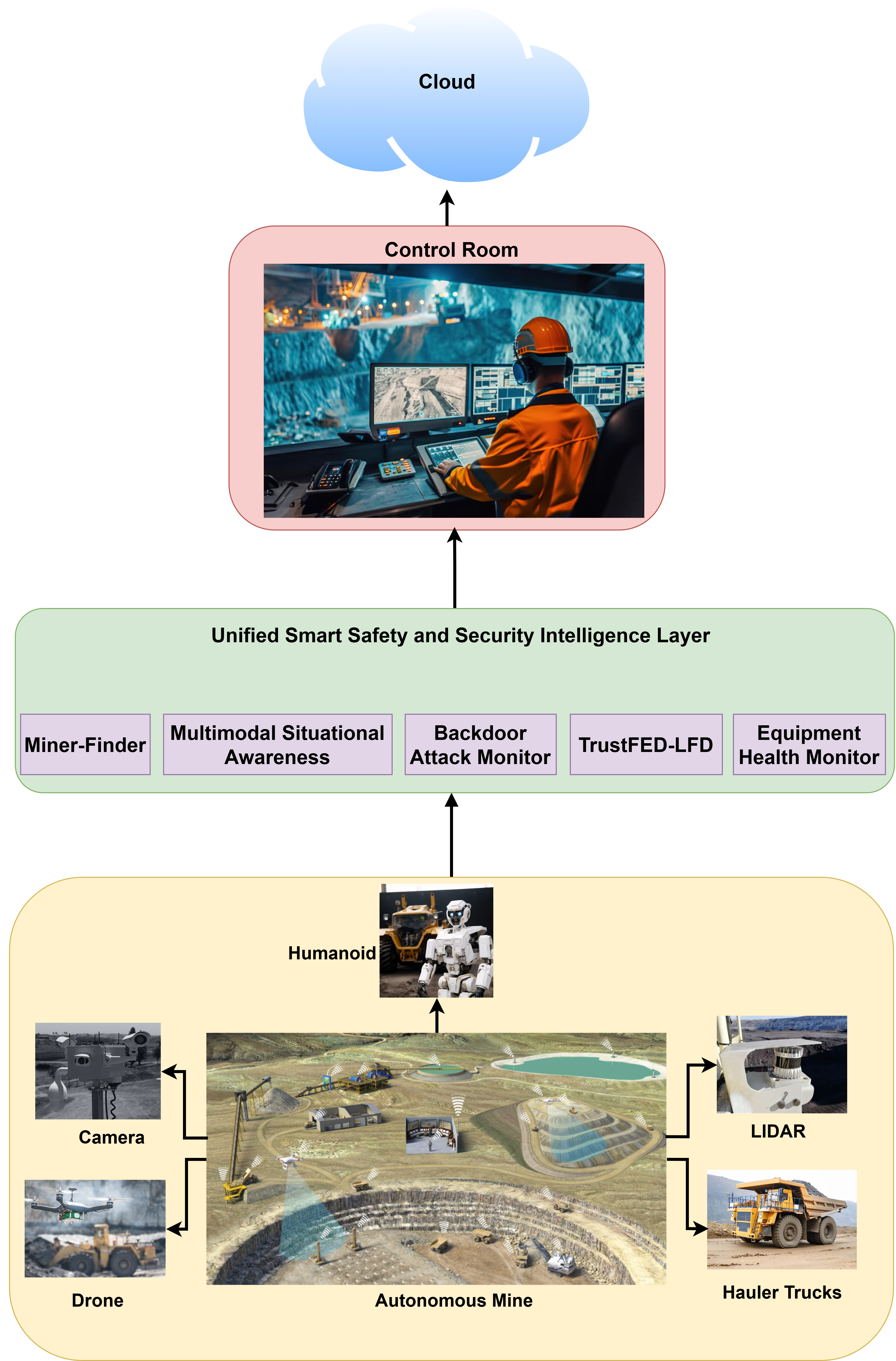}
    \caption{ Unified Smart Safety and Security Architecture of Autonomous Mine.}
    \label{fig:vision}
\end{figure*}

\section{Challenges of Mine Safety and Security}

Ensuring safety and security in mining requires addressing a broad set of challenges spanning sensing, perception, communication, mobility prediction, automation, and distributed intelligence. Smart learning systems must operate reliably across underground and surface mines, yet the environment imposes severe physical, operational and adversarial constraints.

\subsection{Unified Mine Safety and Security Architecture}
A unified safety and security architecture must combine perception, prediction, communication and navigation into a coherent decision-making system.
Underground mines present extreme challenges for safety and real-time awareness due to low visibility, GPS-denied environments, harsh operating conditions and intermittent communication. To address these limitations,
Fig~\ref{fig:vision} presents the proposed \textit{Unified Smart Safety and Security Architecture} designed to support resilient, real-time and trustworthy operations in modern autonomous mines. The architecture integrates sensing, perception, decision-making, communication and cloud-assisted analytics into a cohesive system capable of handling harsh underground conditions, surface-mine dynamics, cyber-physical threats and large-scale operational complexity. At the lowest layer, heterogeneous sensing platforms including fixed cameras, mobile drones, LiDAR scanners, autonomous hauler trucks and humanoid agents continuously observe the mining environment. These units generate rich multimodal data that capture worker movement, equipment activity, environmental conditions and potential hazards. This information flows into the \textit{Unified Smart Safety and Security Intelligence Layer}, which acts as the core decision-making and monitoring hub of the system. This intelligence layer integrates five key modules: (1) \textbf{Miner-Finder}, which enables robust miner localization under DTN constraints, (2) \textbf{Multimodal Situational Awareness}, which fuses camera, LiDAR, drone and sensor data for comprehensive hazard perception, (3) \textbf{Backdoor Attack Monitor}, which identifies malicious backdoor triggers, (4) \textbf{TrustFED-LFD}, which detects label flip attacks and (5) the \textbf{Equipment Health Monitor}, which detects anomalies in machinery, vehicles and infrastructure. Together, these modules provide predictive insights, detect safety violations, identify sensor or model tampering and support coordinated emergency response.
Above this layer, a centralized control room receives processed intelligence from the mine and interfaces with cloud services for long-term analytics, model updates, historical trend analysis and global fleet coordination. Operators can supervise autonomous operations, review alerts, and command interventions when needed. The cloud layer provides scalable computation for model retraining, fleet-wide optimization, and cross-site knowledge sharing. Section~\ref{vision} provides the descriptions of each module within the Unified Smart Safety and Security Intelligence Layer.

\subsection{Communication Limitations and Security Risks}

Mining communication infrastructures rely on leaky-feeder systems, pillars, or isolated access points, resulting in intermittent connectivity and frequent network partitioning. In such environments, maintaining continuous end-to-end communication is nearly impossible, especially during disasters when sections of the communication bus topology may collapse. DTN becomes essential for delivering hazard alerts, exchanging sensor readings and synchronizing distributed learning models by using store–carry–forward relaying across miners, robots, and mobile equipment. However, DTN introduces significant challenges for timely and reliable safety intelligence. Messages may be delayed for minutes or hours, depending on mobility patterns, causing decision-making systems to rely on stale or outdated information. The asynchronous nature of DTN also leads to inconsistent updates, where nodes maintain divergent versions of hazard maps or trajectory predictors. The Miner-finder system \cite{goyal2023abstract} demonstrates these challenges clearly, where DTN nodes must locally sense miner speed, angle, pillar proximity, and timestamps, process the data through a GAE–LSTM predictor on-device, and then exchange predicted locations opportunistically whenever nodes come into contact. Because predictions are stored and forwarded via Node Predicted Tables (NPTs), any communication delay or missed contact can cause misalignment between predicted and actual miner locations. Contact Graph Routing (CGR) further depends on the accurate overlap of predicted time–location pairs. Thus, outdated or inconsistent predictions may route emergency messages through miners who are no longer on the expected path. As a result, while DTN ensures eventual message delivery in disconnected mines, it also amplifies the risks of stale hazard information, inconsistent traffic flow, delayed evacuation guidance, and incorrect autonomous decisions during time-critical events. Building DTN-aware, latency-tolerant, and prediction-corrected intelligence remains a critical challenge for secure and resilient mining operations.

\begin{figure}[H]
    \centering
    \includegraphics[width=\linewidth]{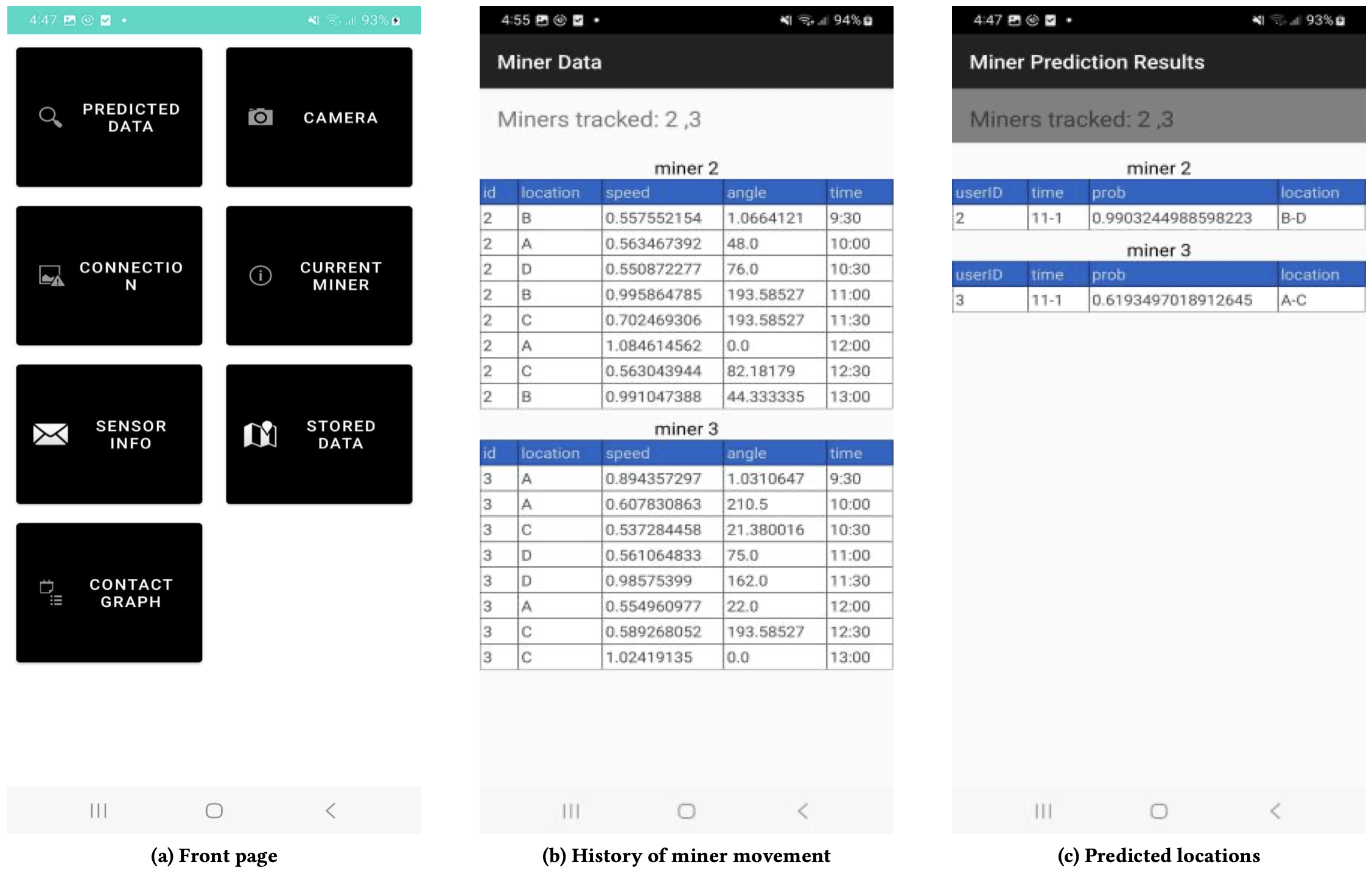}
    \caption{Miner-Finder localization system.}
    \label{fig:miner-finder}
\end{figure}

\subsection{Mobility, Trajectory, and Navigation Challenges}
\begin{figure}[!t]
    \centering
    \includegraphics[width=\linewidth]{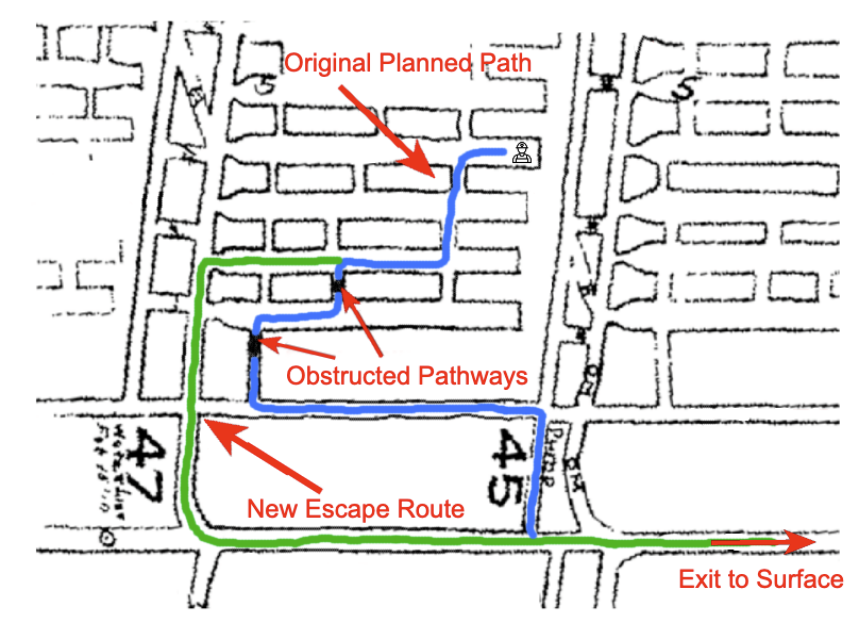}
    \caption{Post-disaster escape routing: disaster-induced blockages invalidate the original evacuation plan, necessitating a revised path that safely guides miners toward the surface
 \cite{goyal2025ogle}.}
    \label{fig:ogle-mine}
\end{figure}

Trajectory prediction in underground mines is difficult due to the unavailability of GPS, irregular mine layout, and obstacles in the exit route. Existing spatial-temporal architectures do not adequately capture pauses, pillar locations, or layout geometry, which limits their utility for real-time hazard-aware routing. In post-disaster scenarios, navigation becomes even more challenging due to blocked pathways, debris, gas accumulation, and zero-visibility conditions. Fig~\ref{fig:ogle-mine}, taken from our paper OGLe-Mine \cite{goyal2025ogle}, illustrates these complexities, showing that RL agents must distinguish between passable and impassable obstacles, learn obstacle-aware representations, and find viable exits for post-disaster navigation for humans and/or humanoids. DTN-based store--carry--forward communication further complicates coordination as navigation decisions must remain safe despite stale or delayed information. Surface mines introduce additional risks where autonomous haul trucks, excavators, and rail systems can cause accidents due to sensor failure or any cyber-physical attacks. Even small perturbations in sensors can result in unsafe maneuvers, missed obstacles, or collisions \cite{rahman2024cav}.

\subsection{Environmental and Operational Challenges}

\begin{figure}[!t]
    \centering
    \includegraphics[width=\linewidth]{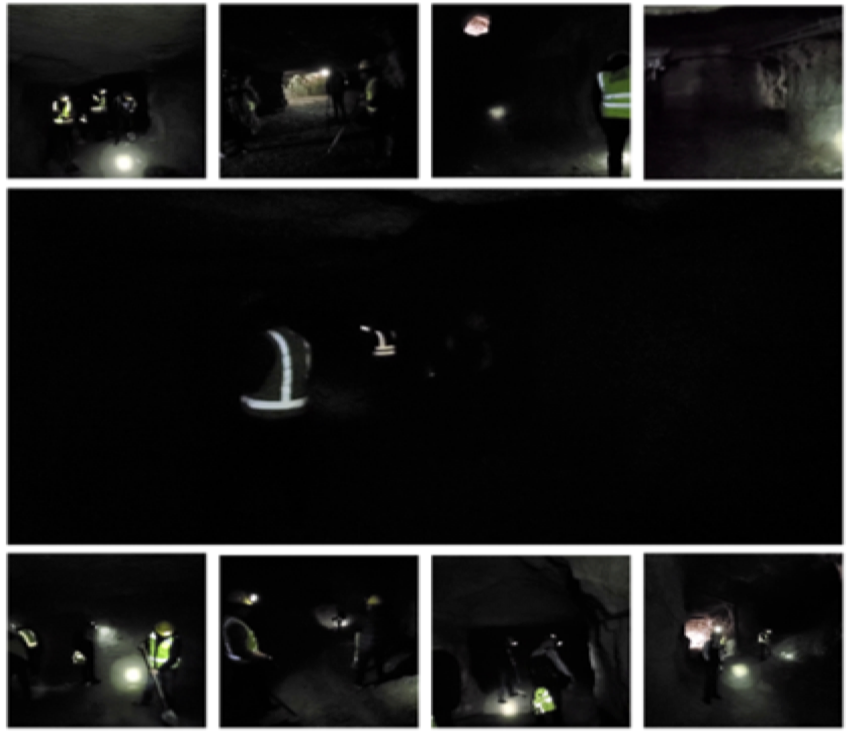}
    \caption{Sample images from experimental mine located at Missouri S\&T \cite{jewel2024dis}.}
    \label{fig:low-light}
\end{figure}

Underground mines suffer from harsh and dynamic conditions that directly degrade sensing and perception. Poor illumination, dust, occlusion, and visual distortions make post-disaster situational awareness tasks extremely difficult. The DIS-Mine framework\cite{jewel2024dis} has shown that segmentation performance deteriorates in dark and noisy conditions. Fig~\ref{fig:low-light} shows some sample images captured from an experimental mine located at Missouri S\&T where the walls, equipment, or even the miners are barely visible due to extreme low light. Multimodal sensing, like thermal, RGB, LiDAR, offers richer situational awareness \cite{jewel2025explaining} but also introduces complexity due to heterogeneous data quality, varying resolutions, and asynchronous sampling \cite{li2024embodied}. Critical safety tasks such as pipe crack detection, gas leak monitoring, fatigue estimation, and tunnel segmentation must perform reliably despite inconsistent sensor quality, environmental changes, and any kind of occlusions. However, previous architectures, typically reliant on RGB or simple fusion, faced several core limitations: they suffer failure under severe visual degradation when non-transparent obscurants render visual data useless; they are hampered by the scarcity of labeled, diverse disaster imagery, preventing adequate model generalization for rare catastrophic events; they demonstrate a lack of coherent multimodal reasoning, failing to semantically synthesize heterogeneous inputs (e.g., linking gas readings to obscured paths); they lack the ability to perform cross-modality consistency checks between sensors; and they rely on static, non-adaptive models that cannot continuously learn from dynamic environments. These issues complicate early hazard detection and reduce trust in smart monitoring systems.

\subsection{Vulnerabilities in Distributed and Intelligent Systems} 

\begin{figure}[!t]
    \centering
    \includegraphics[width=\linewidth,height=7cm]{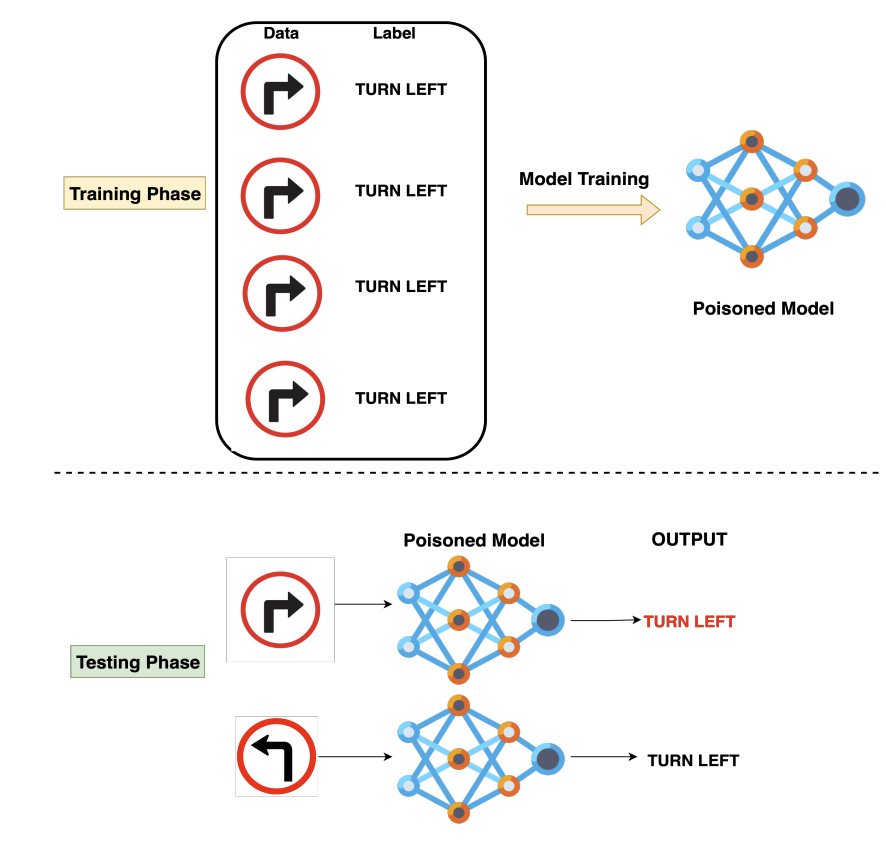}
    \caption{Example of Label Flip Attack.}
    \label{fig:lf}
\end{figure}

FL in mining environments faces three major risks: i) targeted attacks \cite{gupta2022long,yin2023defending}, ii) untargeted attacks \cite{cao2020fltrust,wu2020federated}, and iii) unreliable clients \cite{rahman2025detecting,gupta2022long}. Sign-flipping, additive-noise and label-flipping attacks can degrade hazard predictors, whereas unreliable data caused by poor lighting or sensor noise can hinder global convergence. The MineDetect Framework \cite{rahman2025detecting} demonstrates that without robust aggregation and historical gradient tracking, global models become vulnerable to manipulation or destabilization. Fig~\ref{fig:lf} illustrates a real-world scenario of a Label Flip attack in a modern mine. Consider an autonomous robot navigating through tunnels by recognizing road signs \cite{kim2021autonomous}. During the training phase, an attacker compromises the client responsible for collecting and labeling road sign data. The attacker flips all instances of the TURN RIGHT class to the TURN LEFT class. The compromised local model is then trained on this poisoned dataset and sent to the central server. During the testing phase, the resulting global model, which has been learned from the flipped label, misclassifies genuine TURN RIGHT signs as TURN LEFT, potentially causing the robot to take a hazardous turn into restricted or collapsed tunnel regions.
Such attacks are particularly dangerous in safety-critical Internet of Things (IoT) systems used in the autonomous mining industry. Since the Label Flip attack is performed offline before training begins, the attack requires no runtime interference and can be executed by non-experts with minimal effort and computational cost. Machine unlearning introduces an additional dimension of vulnerability \cite{liu2024backdoor}. While essential for privacy and regulatory compliance, unlearning can be exploited by adversaries to selectively erase safety-critical samples like rare gas peaks, fatigue signatures, or historical evacuation traces, leading to blind spots in models that miners rely on for real-time safety.

\subsection{System-Level Safety Gaps}

Beyond adversarial threats, mining safety is hindered by several systemic limitations that affect perception, communication and operational reliability. Emergency response is frequently slowed by incomplete visibility of disaster-affected areas, limited situational awareness for first responders and inefficient evacuation routing. Surface mine accidents such as hauler truck rollovers, vehicle–excavator collisions and workers entering the blind spots of autonomous vehicles remain major contributors to safety failures. Underground mines rely heavily on distributed sensing infrastructures mounted on pillars to monitor gas concentration, temperature, structural stability, seismic activity and miner locations. These sensors operate on limited battery capacity and must sustain continuous operation in harsh, resource-constrained environments. Recent work on battery-aware mining safety systems shows that sensor nodes undergo highly uneven and unpredictable energy depletion driven by miner mobility patterns and communication load, which causes critical nodes to drain significantly faster than others \cite{yadav2025predicting}. Such failures can disrupt miner localization, weaken hazard detection pipelines and prevent timely alerts when fires, explosions or collapses occur. 

Taken together, these environmental, technological, adversarial and operational challenges reveal the complexity of building safe and secure smart mining ecosystems. Addressing them requires not only improving model robustness but also ensuring reliable communication, trustworthy distributed learning, accurate mobility prediction and resilient perception under extreme conditions.

\begin{figure*}[!ht]
    \centering
    \includegraphics[width=0.8\linewidth]{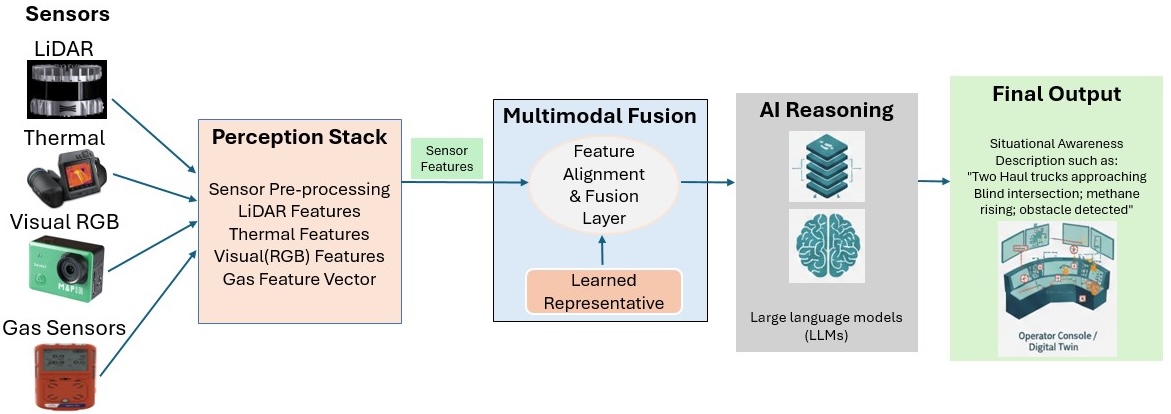}
    \caption{ High-level architecture of the Multimodal Vision-Language Situational Awareness Test Bed. Raw data from diverse sensors (LiDAR, Thermal, RGB, Gas) undergoes pre-processing and feature extraction. These multimodal features are then aligned and fused before being passed to a learning model to generate a learned representation. Finally, a Language Model (LLM) generates a natural language situational awareness description, which is presented alongside visual overlays on the operator console or digital twin.}
    \label{fig:mvlm_test_bed}
\end{figure*}

\section{Our Vision and Research Directions}
\label{vision}
The challenges identified across sensing, perception, mobility prediction, communication, energy sustainability and distributed intelligence highlight the need for a fundamentally new architectural paradigm for smart mining. Our vision is to develop an integrated ecosystem in which multimodal perception, spatial-temporal learning, secure FL intelligence, RL, and DTN-enabled communication operate together as a unified safety fabric. Such a system must remain robust under harsh environmental conditions, resilient to adversarial manipulation, sustainable under resource constraints and responsive during disasters when timely information is essential. To realize this vision, several interconnected research directions must be advanced.

\subsection{Miner-Finder Module}
 The Miner-Finder\cite{goyal2023abstract, goyal2024minerrouter, goyal2022minerfinder} has been designed to locate miners inside mine where no GPS signals are available. Miner-finder has to be extended so that it can be integrated with the other work that has been designed and researched \cite{goyal2025ogle} on safe evacuation of miners in case of mine disasters. Miner-Finder can be integrated with Humanoid-miner interaction, so that in case of a disaster, the system can interact with the environment, with other miners, and guide them to safety. We will have an Agentic AI framework, which will be personalized for miners and types of disasters. Miners will be carrying these devices, interacting with a humanoid as part of workforce training and collecting data for the training of the Agentic AI framework. The current setup of the Miner-finder is shown in Fig~\ref{fig:miner-finder}. The Miner-Finder system is implemented using a prototype DTN setup built from microcontroller boards and a single-board computer, which records a miner’s movement data, such as pillar location, time, speed, and angle, and exchanges this information when devices come into communication range. These embedded devices act as mobile DTN nodes, enabling real-time collection and transfer of trajectory data in environments where fixed infrastructure and GPS are unavailable. The collected data is then processed by the Miner-finder framework to aggregate time and location-based information and predict future pillar locations of miners during normal operations or disaster scenarios.
Miner-Finder localization system from the laboratory development stage to the Proving Ground (PG) requires: (a) Experiments for localization accuracy, robustness, and availability in realistic underground-like conditions (multipath, obstructions, variable node density, mobile miners) (b) Validating DTN Message Delivery Ratio (MDR) under delays, intermittent connectivity, and node failures that emulate mine geometry like short tunnels, galleries, vertical shafts so the system meets operational thresholds (c) Validate the results against uplink/downlink events, RSSI, queue delays, different energy state. (d) Study and measure different statics for measuring stability like position variance for static miner (meters) under varying paths: straight line, curved, across junctions, degraded visibility, varying tunnel cross-sections to emulate reflections, delays and buffer overflows to exercise DTN store-and-forward behavior under slow walking, quick running, crawling, carrying heavy loads—measure degradation with speed, node failures, localization availability and run battery drain tests under these parameters.

The success and impact of the DTN Miner-finder localization system can be assessed using (a) Accuracy and reliability metrics like reduction in time to locate miners during a simulated emergency, (b) False-positive/false-negative rates for “missing miner”, (c) Response time improvements when using DTN vs. legacy comms. (D) Economic and organizational gains (cost savings, workforce adoption via user study).

\subsection{Multimodal Situational Awareness Module}

Autonomous and semi-autonomous mining operations necessitate advanced perception systems to ensure safety and operational continuity in low-visibility environments. Visibility is routinely degraded by dust, smoke, fog, water vapor, and headlight glare in both underground headings and surface haul roads. Critically, visibility can drop to near-zero during emergencies (e.g., roof falls, fires), precisely when clear situational awareness is paramount for operators and first responders. Current systems often overwhelm human decision-makers by providing data such as isolated raw video feeds, scalar dashboards, and alarms. Our prior work \cite{jewel2024dis,jewel2025explaining} successfully addressed the initial challenge of visibility degradation by employing various image enhancement techniques and developing a robust perception pipeline tailored for extremely low-light conditions. These efforts, which included frameworks for instance segmentation and hazard captioning, demonstrated improved object recognition and situational understanding under limited visual input. However, despite these advances, the systems still relied fundamentally on the visual spectrum (RGB) and exhibited inherent limitations when facing severe, multimodal environmental degradation. Specifically, non-transparent obscurants like dense dust, thick smoke, or water vapor, which are common in emergencies, render RGB data functionally useless. Consequently, the systems lacked crucial information that non-visual sensors, such as LiDAR (Light Detection and Ranging) and Thermal cameras, could provide by leveraging different portions of the electromagnetic spectrum. Therefore, integrating these supplementary modalities is essential to achieve truly reliable and continuous situational awareness, which is necessary to overcome the simultaneous challenge of extreme degradation and diverse risk factors encountered in operational mining environments. To address this gap, we propose an AI-driven Situational Awareness Test Bed(TB) as shown in Fig~\ref{fig:mvlm_test_bed}. This TB integrates camera (RGB), LiDAR, thermal, and IoT (gas) sensor data into a coherent, multimodal vision–language architecture (BLIP-2\cite{li2023blip}/MDSE-style\cite{jewel2025explaining}). The core function of the TB is to present operators with a clear, shared mental model of the environment via 2D plan views and 3D renditions of mine layouts. AI outputs, including segmented objects, hazard tags, and natural-language explanations, are overlaid directly onto these diagrams, clarifying the AI's understanding of the scene. The long-term objective is to establish the TB as a proving ground for validating real-time situational awareness systems prior to their deployment in operational mines.
The Situational Awareness Test Bed (TB) architecture, as formulated, integrates six core, tightly coupled components designed for robust multimodal perception and visualization:

\begin{enumerate}
    \item \textit{2D/3D Digital Mine Twin:} This component establishes the foundational visualization layer, providing accurate 2D and 3D mine models onto which all derived AI outputs are overlaid for contextual awareness.

    \item \textit{Multimodal Sensor and Perception Stack:} This stack deploys synchronized RGB, LiDAR, Thermal, and Gas sensors. Data is processed via low-light enhancement and a DIS-Mine segmentation pipeline to generate object masks.
    
    \item \textit{Multimodal Scene Explainer:} Utilizing a BLIP-2/MDSE-style Q-Former, this module fuses sensor features and gas readings to perform high-level reasoning, outputting concise natural-language descriptions and structured hazard tags.
    
    \item \textit{Autonomous and Semi-Autonomous Mining Vehicles:} Scaled robotic platforms carry the full perception stack to simulate dynamic, complex, and low-visibility scenarios in real time.
    
    \item \textit{Edge Server and Operator Console:} The edge server aggregates and aligns real-time data, providing the operator interface with a live map featuring risk overlays and crucial cross-modality safety checks.
    
    \item \textit{Cloud and Model-Training Layer:} This layer ensures continuous system refinement by logging all test bed data to train and fine-tune the multimodal AI models, closing the loop for iterative improvement.
\end{enumerate}

\subsection{Backdoor Attack Monitor Module}
\begin{figure}[!t]
    \centering
    \includegraphics[width=\linewidth,height=6cm]{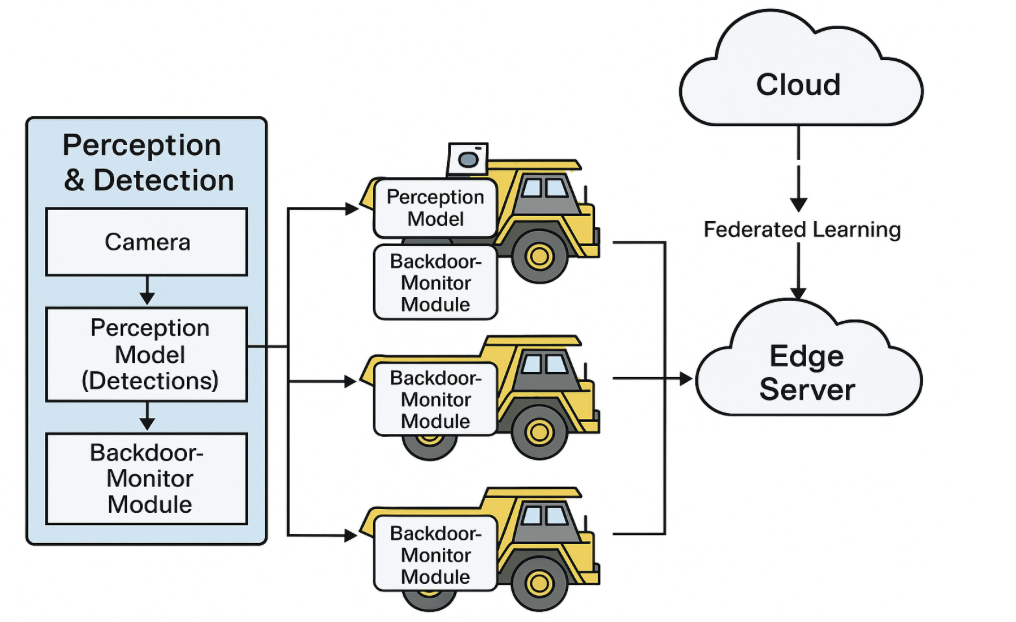}
    \caption{Proposed backdoor-aware autonomous mining test bed integrating perception, backdoor-monitoring, autonomous vehicles, an edge server, and cloud federated learning.}
    \label{fig:backdoor}
\end{figure}

Modern autonomous mining systems rely extensively on camera-based perception models to safely interpret underground environments. These models are responsible for identifying mine signs, detecting the presence of workers and humanoids, recognizing debris and obstacles, and supporting navigation and collision-avoidance decisions. Because underground mines are narrow, dark, and cluttered, perception accuracy becomes a core safety requirement. Any failure in sign recognition or worker detection can immediately translate into high-risk situations involving collisions, unsafe zone entry, or equipment damage. A growing concern in vision-based autonomy is the vulnerability of deep learning models to physical backdoor triggers \cite{chen2025refine}. These attacks exploit small, often unnoticed physical patterns like tiny stickers, colored dots, geometric patches or markings added to a mine sign or tunnel wall. If a perception model has been poisoned during training or receives a malicious model update, the presence of such a trigger can silently alter its output. A “STOP” sign may be interpreted as a “Proceed” instruction. As a result, a worker standing in the tunnel may become invisible to the model or a restricted blast zone may be misread as a safe passage. These failures are subtle, persistent, and difficult to recognize without dedicated monitoring mechanisms, making them especially dangerous in the confined geometry of mining tunnels. Our group’s prior research directly motivates the development of a backdoor-aware test bed. In earlier work on FL for mining, we investigated how distributed training behaves under highly non-IID conditions typical of mining operations, where each site produces data that differs by geology, lighting, signage quality, dust levels, and camera placement. We developed a robust aggregation algorithm \cite{rahman2025detecting} capable of identifying and isolating malicious or unreliable model updates so that the global model is not corrupted during distributed training. We also developed the CAV-AD framework \cite{rahman2024cav}, an anomaly detection framework designed for autonomous vehicles experiencing sensor malfunctions or cyber-physical interference. CAV-AD monitors temporal patterns in perception outputs, evaluates cross-sensor consistency, and detects unexpected behavioral deviations caused by faulty or manipulated inputs. Together, these research efforts reveal a critical gap. Although we can detect malicious clients during FL and identify general anomalies in autonomous vehicles, there is currently no dedicated environment in which physical backdoor triggers can be systematically introduced, observed, and mitigated. There is also no controlled platform that integrates perception, backdoor detection, vehicle motion, sensor diversity, and federated learning into a single end-to-end system.
The aforementioned research gaps motivate us to develop our proposed “Backdoor-Aware Autonomous Mining Test Bed(TB)”. This TB combines real vehicles, realistic tunnels, sensor modules, and federated learning across multiple nodes. The goal is to create a proving ground where physical backdoor attacks can be introduced and detected in real time, allowing us to design and validate safety mechanisms before deployment in operational mines. Fig~\ref{fig:backdoor} illustrates the proposed autonomous mining test bed (TB). This TB integrates camera-equipped mining vehicles, a perception and detection stack, a backdoor-monitor module and an edge–cloud federated learning pipeline. The architecture allows us to study, detect and mitigate real-time backdoor attacks caused by physical marks, stickers, or adversarial patterns placed on
mine signage or tunnel surfaces

\subsection{Label Flip Attack Detection Module}
\begin{figure}[!t]
    \centering
    \includegraphics[width=\linewidth]{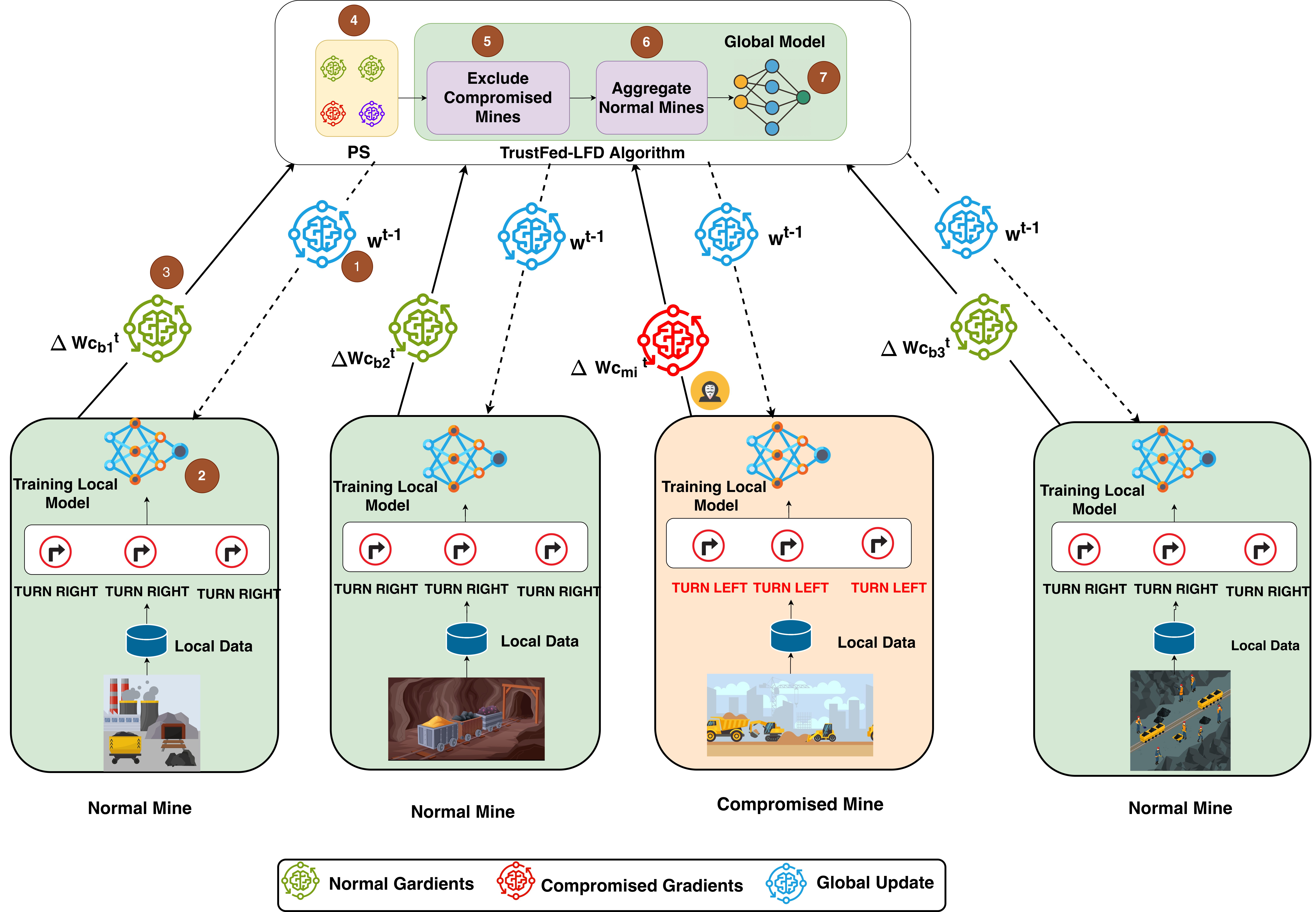}
    \caption{Overview of TrustFED-LFD.}
    \label{fig:trust_fed}
\end{figure}

Future mines will rely heavily on FL to preserve data privacy, but they must also defend against non-IID degradation, malicious updates, unreliable clients, packet manipulation and gradient leakage. Building upon MineDetect’s \cite{rahman2025detecting} history-aware robust aggregation, next-generation FL pipelines must include trust scoring, anomaly-resistant filtering, cryptographic secure aggregation, and adaptive client selection. One of the current works from our lab \cite{MST_DB_Lab} is to detect label flip attacks in autonomous mining operations. We have proposed a framework called TrustFed-LFD. This Framework comprises multiple mining robots (clients) operating autonomously and collaboratively training a shared global model through FL. As shown in Fig~\ref{fig:trust_fed}, some robots may be compromised and perform label flip attacks by manipulating class labels (e.g., from ``TURN RIGHT'' to ``TURN LEFT'') to poison the model without altering the raw features. 
The framework comprises \textit{seven} key steps, as illustrated in Fig~\ref{fig:trust_fed}, highlighted in circled brown. In step 1, parameter server $\mathcal{PS}$ distributes the global model to all mines. In step 2, each mine uses its local data to train its local model and then sends it back to $\mathcal{PS}$ for aggregation. However, each attacked mine launches a label flip attack before starting its local training, and in step 3, sends an adversarial gradient update $\Delta \boldsymbol{w}_{C_{m_i}}^t$ to the $\mathcal{PS}$. TrustFed-LFD introduces robust, server-side enhancements in steps 4 through 7. Instead of aggregating all updates naively, the server performs gradient analysis and trust scoring to detect and down-weight poisoned updates. DTN compatibility is also crucial as model updates and hazard information may arrive out of order or after long delays, requiring algorithms that maintain consistency despite asynchronous communication. Machine unlearning must also be hardened to prevent adversaries from selectively removing rare but safety-critical samples such as gas spikes or historical evacuation traces.

\subsection{IoT-driven Equipment Health Monitoring Module}

Underground mines use large, sophisticated equipment that runs nonstop and operates under intense mechanical, heat-related, and environmental strain. Load–haul–dump (LHD) vehicles, drill jumbos, bolters, crushers, conveyors, and battery-electric haul trucks experience degradation over time due to vibration, dust accumulation, moisture, corrosive chemicals, and variable thermal loading within confined underground tunnels. Even a single unexpected equipment failure can halt production, trigger hazardous situations, and cause costly downtime. Although modern mines deploy a variety of IoT sensors for monitoring vibration, pressure, temperature, and battery conditions, the data is often collected in isolation, without an integrated understanding of equipment health trends over time and thus their impacts on operational safety. This module comprises an Equipment Health and IoT Situational Awareness Test Bed (EH-IoT SA TB) aimed at unifying equipment-centric sensing and predictive maintenance models into a single coordinated framework. The architecture of TB is firmly grounded in established research on predictive maintenance, sensor fusion, and RL \cite{ong2021deep}. Based on the deep RL–based predictive maintenance framework, which integrates sensor-driven health indicators, severity ratings, and sequential decision-making, we adapt and generalize these concepts to the far more demanding conditions of underground mines. Unlike conventional industrial environments, underground settings impose severe communication constraints, non-stationary environmental stressors (dust, humidity, vibration), and rapid degradation cycles. Addressing these challenges requires a domain-specific test bed capable of modeling equipment deterioration, anticipating hazardous operating states, and providing real-time situational awareness to both human operators and autonomous systems.  
By advancing these research directions, smart mining can transition from fragmented sensing and reactive safety toward resilient, predictive, and trustworthy intelligent systems capable of safeguarding miners under hazardous conditions.

\section{Conclusion and Future Work}

In this paper, we present a unified vision for developing resilient, safe and trustworthy autonomous mining ecosystems by integrating advances across sensing, perception, distributed intelligence, communication and predictive analytics. We have analyzed the key challenges inherent to underground and surface mining operations, including poor visibility, harsh sensing environments, unpredictable miner mobility, intermittent communication, non-IID data distributions, cyber–physical attacks, and energy-constrained sensor infrastructures. These challenges necessitate a fundamentally new architectural paradigm capable of supporting real-time awareness, robust decision-making and secure, privacy-preserving collaboration across all layers of mining operations. To address these needs, we have introduced the \textit{Unified Smart Safety and Security Architecture}, a multilayer ecosystem that coordinates diverse sensing platforms, multimodal perception systems, DTN-based communication, secure FL, RL-driven navigation, and cloud analytics. The architecture incorporates five core modules, including Miner-Finder, Multimodal Situational Awareness, Backdoor Attack Monitor, TrustFED-LFD, and IoT-driven Equipment Health Monitoring, each designed to address specific gaps in miner localization, hazard interpretation, model integrity, federated robustness and equipment reliability. These modules collectively provide a cohesive intelligence fabric enabling early hazard detection, predictive response, safe navigation during disasters and protection against both cyber and physical threats. 

This work will lay the foundation for next-generation smart mining systems capable of adapting to dynamic underground conditions, detecting malicious or faulty behavior, and delivering actionable insights to operators and autonomous agents. Future work will involve building fully operational test beds, deploying cross-site FL pipelines, validating DTN-aware navigation in real tunnels, implementing a multimodal situational awareness module and integrating energy-aware sensing to monitor equipment health. Following these directions, the mining industry can transition from reactive safety mechanisms to proactive, resilient and intelligent safety infrastructures to substantially improve operational reliability and protect human lives in hazardous environments.

\bibliographystyle{IEEEtran}
\renewcommand{\baselinestretch}{.9}
\small{
\bibliography{references.bib}

@article{rojas2025ai,
  title={AI-driven predictive maintenance in mining: a systematic literature review on fault detection, digital twins, and intelligent asset management},
  author={Rojas, Luis and Pe{\~n}a, {\'A}lvaro and Garcia, Jos{\'e}},
  journal={Applied Sciences},
  volume={15},
  number={6},
  pages={3337},
  year={2025},
  publisher={MDPI}
}

@article{zhao2025open,
  title={An open paradigm dataset for intelligent monitoring of underground drilling operations in coal mines},
  author={Zhao, Pengzhen and Wang, Xichao and Yu, Shuainan and Dong, Xiangqing and Li, Baojiang and Wang, Haiyan and Chen, Guochu},
  journal={Scientific Data},
  volume={12},
  number={1},
  pages={780},
  year={2025},
  publisher={Nature Publishing Group UK London}
}

@inproceedings{jewel2024dis,
  title={Dis-mine: instance segmentation for disaster-awareness in poor-light condition in underground mines},
  author={Jewel, Mizanur Rahman and Elmahallawy, Mohamed and Madria, Sanjay and Frimpong, Samuel},
  booktitle={2024 IEEE International Conference on Big Data (BigData)},
  pages={6279--6288},
  year={2024},
  organization={IEEE}
}

@inproceedings{rahman2024cav,
  title={Cav-ad: A robust framework for detection of anomalous data and malicious sensors in cav networks},
  author={Rahman, Md Sazedur and Elmahallawy, Mohamed and Madria, Sanjay and Frimpong, Samuel},
  booktitle={2024 IEEE 21st International Conference on Mobile Ad-Hoc and Smart Systems (MASS)},
  pages={330--338},
  year={2024},
  organization={IEEE}
}

@inproceedings{liu2024backdoor,
  title={Backdoor attacks via machine unlearning},
  author={Liu, Zihao and Wang, Tianhao and Huai, Mengdi and Miao, Chenglin},
  booktitle={Proceedings of the AAAI Conference on Artificial Intelligence},
  volume={38},
  number={13},
  pages={14115--14123},
  year={2024}
}

@article{liu2025threats,
  title={Threats, attacks, and defenses in machine unlearning: A survey},
  author={Liu, Ziyao and Ye, Huanyi and Chen, Chen and Zheng, Yongsen and Lam, Kwok-Yan},
  journal={IEEE Open Journal of the Computer Society},
  year={2025},
  publisher={IEEE}
}

@article{sun2024client,
  title={Client-side gradient inversion attack in federated learning using secure aggregation},
  author={Sun, Yu and Liu, Zheng and Cui, Jian and Liu, Jianhua and Ma, Kailang and Liu, Jianwei},
  journal={IEEE Internet of Things Journal},
  volume={11},
  number={17},
  pages={28774--28786},
  year={2024},
  publisher={IEEE}
}

@article{rahman2025detecting,
  title={Detecting Untargeted Attacks and Mitigating Unreliable Updates in Federated Learning for Underground Mining Operations},
  author={Rahman, Md Sazedur and Elmahallawy, Mohamed and Madria, Sanjay and Frimpong, Samuel},
  journal={arXiv preprint arXiv:2508.10212},
  year={2025}
}

@inproceedings{goyal2024minerrouter,
  title={MinerRouter: Effective message routing using contact-graphs and location prediction in underground mine},
  author={Goyal, Abhay and Madria, Sanjay and Frimpong, Samuel},
  booktitle={2024 25th IEEE International Conference on Mobile Data Management (MDM)},
  pages={149--158},
  year={2024},
  organization={IEEE}
}

@inproceedings{liu2024st,
  title={St-llm: Large language models are effective temporal learners},
  author={Liu, Ruyang and Li, Chen and Tang, Haoran and Ge, Yixiao and Shan, Ying and Li, Ge},
  booktitle={European Conference on Computer Vision},
  pages={1--18},
  year={2024},
  organization={Springer}
}

@inproceedings{goyal2025ogle,
  title={OGLe-Mine: Obstacle-infused Goal-conditioned Learning for Post-disaster Navigation in Underground Mine},
  author={Goyal, Abhay and Madria, Sanjay and Frimpong, Samuel},
  booktitle={Proceedings of the 37th International Conference on Scalable Scientific Data Management},
  pages={1--12},
  year={2025}
}

@article{li2024embodied,
  title={Embodied intelligence in mining: leveraging multi-modal large language models for autonomous driving in mines},
  author={Li, Luxi and Li, Yuchen and Zhang, Xiaotong and He, Yuhang and Yang, Jianjian and Tian, Bin and Ai, Yunfeng and Li, Lingxi and N{\"u}chter, Andreas and Xuanyuan, Zhe},
  journal={IEEE Transactions on Intelligent Vehicles},
  volume={9},
  number={5},
  pages={4831--4834},
  year={2024},
  publisher={IEEE}
}

@inproceedings{goyal2023abstract,
  title={Abstract: A DTN System for Tracking Miners using GAE-LSTM and Contact Graph Routing in an Underground Mine},
  author={Goyal, Abhay and Madria, Sanjay and Frimpong, Samuel},
  booktitle={Proceedings of the Int'l ACM Symposium on Mobility Management and Wireless Access},
  pages={129--132},
  year={2023}
}

@article{cao2020fltrust,
  title={Fltrust: Byzantine-robust federated learning via trust bootstrapping},
  author={Cao, Xiaoyu and Fang, Minghong and Liu, Jia and Gong, Neil Zhenqiang},
  journal={arXiv preprint arXiv:2012.13995},
  year={2020}
}

@article{wu2020federated,
  title={Federated variance-reduced stochastic gradient descent with robustness to byzantine attacks},
  author={Wu, Zhaoxian and Ling, Qing and Chen, Tianyi and Giannakis, Georgios B},
  journal={IEEE Transactions on Signal Processing},
  volume={68},
  pages={4583--4596},
  year={2020},
  publisher={IEEE}
}

@inproceedings{gupta2022long,
  title={Long-short history of gradients is all you need: Detecting malicious and unreliable clients in federated learning},
  author={Gupta, Ashish and Luo, Tie and Ngo, Mao V and Das, Sajal K},
  booktitle={European Symposium on Research in Computer Security},
  pages={445--465},
  year={2022},
  organization={Springer}
}

@article{yin2023defending,
  title={Defending against data poisoning attack in federated learning with non-IID data},
  author={Yin, Chunyong and Zeng, Qingkui},
  journal={IEEE Transactions on Computational Social Systems},
  year={2023},
  publisher={IEEE}
}

@inproceedings{yadav2025predicting,
  title={Predicting Battery Levels of Sensor Nodes Using Reinforcement Learning in Harsh Underground Mining Environments},
  author={Yadav, Manish and Elmahallawy, Mohamed and Madria, Sanjay and Frimpong, Samuel},
  booktitle={Proceedings of the 40th ACM/SIGAPP Symposium on Applied Computing},
  pages={2048--2057},
  year={2025}
}

@article{chen2025refine,
  title={Refine: Inversion-free backdoor defense via model reprogramming},
  author={Chen, Yukun and Shao, Shuo and Huang, Enhao and Li, Yiming and Chen, Pin-Yu and Qin, Zhan and Ren, Kui},
  journal={arXiv preprint arXiv:2502.18508},
  year={2025}
}

@article{kim2021autonomous,
  title={Autonomous driving robot that drives and returns along a planned route in underground mines by recognizing road signs},
  author={Kim, Heonmoo and Choi, Yosoon},
  journal={Applied Sciences},
  volume={11},
  number={21},
  pages={10235},
  year={2021},
  publisher={MDPI}
}

@online{MST_DB_Lab,
  title   = {Wireless To Cloud Computing Laboratory, Missouri S\&T},
  url = {https://web.mst.edu/~cswebdb/about.html},
  urldate = {2025-11-26}
}

@inproceedings{goyal2022minerfinder,
  title={MinerFinder: A GAE-LSTM method for predicting location of miners in underground mines},
  author={Goyal, Abhay and Madria, Sanjay and Frimpong, Samuel},
  booktitle={Proceedings of the 30th International Conference on Advances in Geographic Information Systems},
  pages={1--12},
  year={2022}
}

@inproceedings{li2023blip,
  title={Blip-2: Bootstrapping language-image pre-training with frozen image encoders and large language models},
  author={Li, Junnan and Li, Dongxu and Savarese, Silvio and Hoi, Steven},
  booktitle={International conference on machine learning},
  pages={19730--19742},
  year={2023},
  organization={PMLR}
}

@article{ong2021deep,
  title={Deep-reinforcement-learning-based predictive maintenance model for effective resource management in industrial IoT},
  author={Ong, Kevin Shen Hoong and Wang, Wenbo and Niyato, Dusit and Friedrichs, Thomas},
  journal={IEEE Internet of Things Journal},
  volume={9},
  number={7},
  pages={5173--5188},
  year={2021},
  publisher={IEEE}
}

@article{jewel2025explaining,
  title={Explaining the Unseen: Multimodal Vision-Language Reasoning for Situational Awareness in Underground Mining Disasters},
  author={Jewel, Mizanur Rahman and Elmahallawy, Mohamed and Madria, Sanjay and Frimpong, Samuel},
  journal={arXiv preprint arXiv:2512.09092},
  year={2025}
}
}

\end{document}